\documentclass[12pt,preprint]{aastex}
\input epsf.tex
\begin{document}
\title{Type Ia Supernovae and High Velocity White Dwarfs}
\author{Brad M.\ S.\ Hansen }
\affil{Hubble Fellow, \\ Div. of Astronomy, 8971 Math Sciences, UCLA, \\ 405 Hilgard Street, Los Angeles, CA, 90095; \\ hansen@astro.ucla.edu}

\begin{abstract}
I examine the hypothesis that many of the high velocity white dwarfs observed by Oppenheimer et al. (2001)
are the remnants of donor stars from binaries that produced type~Ia supernovae via the `single degenerate'
channel. If this channel is a significant contributor to the Galactic SN~Ia supernova rate,
then the local density of such remnants with $V_{\perp}>100 \,{\rm km.s^{-1}}$ could be as high as $2 \times 10^{-4} {\rm pc^{-3}}$,
comparable to the densities found by Oppenheimer et al.
This white dwarf population differs from others in that it is composed exclusively of single stars. 
\end{abstract}


\section{Introduction}

The notion that white dwarfs may play a part in the Galactic dark matter has been actively
investigated in the last few years, motivated primarily by the microlensing observations
(Alcock et al. 1997, 2000). Recently there have been claims that this population has
been detected directly (Oppenheimer et al. 2001). However, neither of these claims is
without detractors. The microlensing optical depth to the LMC may also receive a significant contribution
from LMC-associated populations as well (Sahu 1994; Zhao \& Evans 2000; Alcock et al. 2001), while
the high velocity white dwarf sample may possess a thick disk component (Reid, Sahu \& Hawley 2001;
Reyle, Robin \& Creze  2001; Koopmans \& Blandford 2001; Hansen 2001).

In this paper I wish to investigate a potential source of high velocity white dwarfs different
from the usual population; one which may appear in the searches for halo white dwarfs. The proposed
population results from the formation of white dwarfs from stars in the range $1.3-3\, \rm M_{\odot}$
which act as the donor stars in close mass-transfer binaries in which the accretor is a white
dwarf. Observationally, these binaries are thought to be the origin of the Super-Soft X-Ray
sources. These binaries may also contribute to the type~Ia supernova rate in spiral galaxies.
If so, the explosion of the accretor will release the donor star with its pre-supernova
orbital velocity. Thus, the white dwarf remnants of these donors (hereafter DR, or `donor remnants') will be a population
of high velocity white dwarfs. 
 It is the origin and implications of this population that we
investigate below.

\section{The Single Degenerate Supernova Scenario}
\label{Pathways}

The working hypothesis of this paper is that the observed Type~Ia supernovae are the
result of the so-called `single degenerate' scenario. In this instance, the pre-supernova
white dwarf accretes material from a close, non-degenerate binary companion. This is only
 one of the currently viable models for Type~Ia supernovae (e.g., Branch et al. 1995). Indeed, it
cannot be the only one as it is restricted to younger stellar populations (see below).
However, if most close Super-Soft X-Ray sources lead to
Type~Ia supernovae, they will produce a rate comparable to the estimated Galactic rate (Kahabka \& van den Heuvel 1997; Kahabka 1999).

In order for the white dwarf to grow the accretion rate must meet several severe constraints (Rappaport et al. 1994;
 Kahabka \& van den Heuvel 1997).
First, the accretion rate must be high enough so that steady burning is possible, otherwise 
nova eruptions remove the accumulated mass and the white dwarf never grows (and may even experience
 a net loss of mass). On the other hand, the accretion rate must be low enough that the accretor
envelope does not swell to the point that it engulfs the companion, which would result in a
common-envelope episode, effectively ending the mass transfer.

Following the Rappaport et al. (1994) model, we find that the accretion rate is limited by
\begin{eqnarray}
\dot{M} & < & 8 \times 10^{-7} {\rm M_{\odot} yr^{-1}} \left( M_{\rm wd} - 0.5 {\rm M_{\odot}} \right) \\
\dot{M} & > & 1.3 \times 10^{-7} {\rm M_{\odot}} M_{\rm wd}^{3.6}
\end{eqnarray}

For a donor that is only slightly evolved, so that the
envelope still has a significant radiative component, the mass transfer
occurs on the thermal time of the envelope (if the donor mass is larger
than the accretor mass). Thus,
we approximate the accretion rate as 
\begin{equation}
\dot{M} = 0.5 \left[ M_{\rm d}(0) - M_{\rm wd}(0) \right]/\tau_{\rm KH}
\end{equation}
where $\tau_{\rm KH} = 3.1 \times 10^{7} {\rm yrs} M_{\rm d}^{-2}$ is the Kelvin-Helmholtz time. 
Thus, to meet the requirements for stable mass transfer, we require $1.3 {\rm M_{\odot}}< M_d < 3.3 {\rm M_{\odot}}$.
When the type~Ia supernova occurs, the remnant of the donor will emerge with the
pre-supernova orbital velocity. The fact that $M_{\rm d}>1.3 {\rm M_{\odot}}$ means that {\em all}
the kicked stars end up as white dwarfs. Even after taking into account the significant mass loss,
Langer et al. (2000) find that all the donor remnants have $M > 1 {\rm M_{\odot}}$ (see their
Figure~18).
 Thus, this scenario produces a high velocity population
composed exclusively of white dwarfs.

The velocity distribution depends on the separation of the binaries when the explosion
occurs. The population synthesis of Super-Soft Sources by Rappaport et al. (1994)
reveals binaries with orbital periods in the range 0.3 -- 2 days. This means orbital
velocities of $\sim 100-200 \, {\rm km.s^{-1}}$, depending on the mass ratio of donor and
white dwarf at ignition. This is lower than the velocities found by  Canal, Mendez \& Ruiz-Lapuente (2001) 
because this is the absolute space velocity of the donor, rather than the velocity relative to the
supernova remnant centre (a factor $M_{\rm wd}/(M_{\rm wd}+M_{\rm donor})$ smaller).

\section{The Galactic Population}

To determine the importance of this DR population to high velocity white dwarf studies, we
need to know both what the overall birthrate is and also how the kick velocities modify
the original thin disk kinematics.

Estimates of the local type Ia supernova rate (Tammann, Loffler \& Schroder 1994; van den Bergh
\& McClure 1994) yield $\sim 4 \times 10^{-3} {\rm yr^{-1}}$ for the Galaxy, suggesting a total
population of $\sim 4 \times 10^7$ remnants in the Galaxy. This is an estimate of the {\em current
rate}. One may also estimate a rate based on the total amount of Iron in the Galaxy ($\sim 10^8 M_{\odot}$
assuming solar proportions and $9 \times 10^{10} \rm M_{\odot}$ in the Galactic disk). It depends
on what one assumes for the average output per supernova and the relative contribution of
core collapse and thermonuclear supernovae. Low estimates ($\sim 9 \times 10^7$ type~Ia) are obtained using the results of
Arnett, Schramm \& Truran (1989) ($\sim 0.7 \rm M_{\odot}$ per event and $\sim 50\%$ of the global
Iron from type~II supernovae) while high estimates ($\sim 2.8 \times 10^8$ type~Ia) are obtained using
the numbers from Qian \& Wasserburg (2002) ($\sim 0.3 \rm M_{\odot}$ per event and $\sim 1/3$ from type~II). So,
we will consider a range for the global numbers of remnants born ranging from $4 \times 10^7$ to $2.8 \times 10^8$.

There is also no quarantee that every supernova~Ia arises from the kinds of progenitor system assumed
here. Branch et al. (1995) describe the Super-Soft sources as a promising candidate for the progenitor
populations in young stellar systems such as spiral galaxies, but not for elliptical galaxies (in which
 the $>1.3 \rm M_{\odot}$ objects should have evolved to white dwarfs by now). Hillebrandt \& Niemeyer (2000)
favour the single degenerate scenario above the double degenerate or merging white dwarf scenario.
 Whether a realistic
birthrate can be achieved depends on the uncertain details of population synthesis studies (Yungelson
et al. 1996; Di Stefano \& Nelson 1996; Kahabka \& van den Heuvel 1997). 
I will subsume this uncertainty
into the general rate uncertainty described above.

To describe the effect of the kick velocity, I simulate the DR population by applying randomly oriented
kicks to a thin disk progenitor system.
I use the simple Galactic model of Paczynski (1990), with bulge, halo and disk potentials
\begin{eqnarray}
\Phi_{\rm j} & = & G M_{\rm j}\left( R^2 + \left[ a_{\rm j} + \left( z^2 + b_{\rm j}^2\right)^{1/2} \right]^2 \right)^{-1/2} \\
\Phi_{\rm h} & = & \frac{G M_{\rm h}}{r_{\rm h}} \left[ \frac{1}{2} \ln \left( 1 + \frac{r^2}{r_{\rm h}^2} \right) + \frac{r_{\rm h}}{r}
\tan^{-1} \left( \frac{r}{r_{\rm h}} \right) \right] 
\end{eqnarray}
where $\rm j=b,d$,$M_{\rm b}=1.12 \times 10^{10} \rm M_{\odot}$, $M_{\rm d} = 8.78 \times 10^{10} \rm M_{\odot}$, $M_{\rm h}=5 \times 10^{10} \rm M_{\odot}$,
$a_{\rm b}=0$,$a_{\rm d}=4.2$~kpc, $b_{\rm b}=0.277$~kpc, $b_{\rm d}=0.198$~kpc and $r_{\rm h}=6$~kpc.
I assume stars are born from exponential density distributions in both R and z, with scale lengths, 2.8~kpc and 0.35~kpc
respectively.
I assume the kick velocities are evenly distributed between $100 \rm km.s^{-1}$ and $200 \rm km.s^{-1}$, and randomly
oriented. The epoch of the supernova event is distributed evenly over the last 10~Gyr. The resulting DR proper
motion distribution is shown in Figure~1.

The result of these simulations is a total density near the sun 
\begin{equation}
n_{\odot} \sim 5.3 \times 10^{-4} {\rm pc^{-3}} \left( \frac{N_{\rm tot}}{2.8 \times 10^8} \right)
\end{equation}
where $N_{\rm tot}$ is the total number of type~Ia supernovae that have exploded in the
Galaxy over the last Hubble time. To compare directly with the estimated local density 
from Oppenheimer et al. (2001), we must take only that fraction ($\sim 0.34$) which
passes the proper motion cut $V_{\perp} > 100 \rm km.s^{-1}$ in the solar neighbourhood. 
Thus, the range of possible densities in the Oppenheimer sample, depending on the global supernova rate, is
$n_{\odot} \sim 2.6 \times 10^{-5} {\rm  pc^{-3}}$--$ 1.8 \times 10^{-4} {\rm pc^{-3}}$. Thus, at the
very least the DR density is  comparable to the expected spheroid contribution\footnote{
This can be easily understood as the estimated total stellar mass in a stellar population per SN~Ia
is $\sim 700 \rm M_{\odot}$ (Ruiz-Lapuente \& Canal 2001), similar to the amount of 
 thin disk stellar mass per spheroid star in the solar neighbourhood.} and the value
at the upper end of the allowed range would represent a major contributor to the density
found by Oppenheimer et al.


\subsection{Telling the difference}

The origin of the Oppenheimer et al. (2001) sample has been a subject of some dispute. How do we determine
whether the scenario described above is important or not?
The individual properties of the DR white dwarfs themselves do not reflect much about their origins. However,
we may potentially test this scenario by considering the unique aspects of the population as a whole. The
most important is that the DR population is composed solely of {\em single} stars. The fact
that the kick velocity is actually the orbital velocity of a disrupted close binary excludes the possibility
that any of the DR have a binary companion. Close binary companions are excluded by the fact that
the progenitor system is a mass-transferring close binary, i.e., there is nowhere one could place
a third body that would be dynamically stable and furthermore not interfere with the mass transfer process.
Similarly, a distant third body bound to the original binary would no longer be bound after the supernova.
This follows from the fact that the third body must have a separation at least twice that of the close binary
(to be stable) and a consequently smaller orbital velocity.

The binary frequency amongst the local white dwarfs is 25\% (Holberg, Oswalt \& Sion 2002). One may also
estimate this by the $V_{\rm max}$-weighted binary contribution to the proper motion-selected sample
of Liebert et al. (1988). Using the weights of Leggett, Ruiz \& Bergeron (1998), we find 8\% of the white
dwarf number density lies in close double degenerate stars, whose absolute magnitudes are too bright to
be from a single C/O white dwarf. Another 15\% is contributed by white dwarfs in wide, common-proper motion binaries
with main sequence stars brighter than the faint end of the white dwarf sample. The fraction may be higher
if there are a lot of faint, low-mass companions, but the estimate of 23-25\% is a conservative limit.
 Thus, the acquisition of accurate trigonometric parallaxes for the Oppenheimer sample
should allow us to determine the fraction with anomalously bright luminosities for their colours, i.e.,
unresolved double degenerate binaries\footnote{In principle, single Helium core white dwarfs could
also show this signature. However, Langer et al. (2000) find that all supernova binaries leave donors
with $M>1 \rm M_{\odot}$, which should produce C/O core white dwarfs.}. Further searches for common proper motion companions (as bright
or brighter than the white dwarfs themselves) will constrain the wide binary fraction. With sample
sizes of 100 or more objects, this hypothesis is eminently testable with current capabilities\footnote{We stress, however, that the
binary fraction quoted refers to the $V_{\rm max}$-weighted density, not the simple number count.}.

However, the situation can be complicated if the DR population is mixed with a comparable density of
spheroid white dwarfs or even thick disk white dwarfs (Reid, Sahu \& Hawley 2001; Reyle, Robin \& Creze 2001).
Estimates of the densities from these populations are somewhat uncertain, but we can hope to seperate
them by examining the age distributions. In particular, both these populations are thought to be the
result of old bursts of star formation early in the history of the Galaxy. On the other hand, the
DR population is drawn from a stellar population that is continuously forming stars. Thus, even
if all the other physics is the same, the different star formation histories lead to different
white dwarf luminosity functions.

Consider now a white dwarf population that contains a total density composed equally of thick disk
or spheroid remnants (i.e. a burst population) and DR (a continuously produced
population).
Figure~2 shows the fraction of the white dwarfs in each luminosity bin that belong to the
burst population, i.e. the fainter bins are dominated by the burst dwarfs and the brighter bins by the
DR. This demonstrates how we might disentangle the contributions from the
two populations. If the burst population has a binary fraction of 25\% and the DR population has
no binaries, then the binary frequency of the combined population, for $M_V<14$, is $\sim 1/3 \times 0.25 \sim 8\%$.
Thus, even if the total contributions are similar, the binary fraction amongst the bright white dwarfs is measurably reduced. 

In fact, the Oppenheimer et al. (2001) sample contains a lot of hot white dwarfs, enough so that
 Hansen (2001) has argued that the luminosity function contains a significant fraction of stars
with a star formation history more consistent with the thin disk than an old burst population. If
this is true, then the binary fraction of this sample should be low.
We note that two other scenarios have been suggested to address this problem. Koopmans
\& Blandford (2002) and Davies, King \& Ritter (2001) have also suggested mechanisms by which stars can be
ejected from the thin disk to form an extended population. However, the one described here is the
only one guaranteed to consist entirely of white dwarfs. As such, a control study of binary fractions
in another stellar type (e.g. M dwarfs) with similar kinematics, can potentially also be used to
test this scenario. 

The asymmetric drift of a population can provide some information about the origins. Using the simulations
we compare the
mean tangential velocity at the solar neighbourhood to the circular velocity, and we find an asymmetric
drift $V_{\rm a} \sim V_{\rm circ} - < V_{\phi} > \sim -33 \, {\rm km.s^{-1}}$. So the post-kick kinematics of this population are
 closer to that of a thick disk population than a halo population. Similar drifts are produced by
the Koopmans \& Blandford (2002) \& Davies et al. (2001) scenarios.

\subsection{Microlensing Optical Depth}

It is of interest that the DR population is composed of single stars
only. As such, it circumvents a very popular argument that ascribes the dominant microlensing contribution to
lens populations associated with the Magellanic clouds themselves. The argument in favour of this is that
the binary lenses so far detected (which provide extra information -- namely the lens proper motion) are
located in the clouds (Albrow et al. 1999; Afonso et al. 2000, Alcock et al. 2001). In the case of the DR, no 
such events are even possible.

One can estimate the microlensing optical depth from our simulations
\begin{eqnarray}
\tau & = & \int_0^{D_s} \frac{4 \pi G}{c^2} \rho \frac{D_d (D_s-D_d)}{D_s} dD_d \nonumber \\
 & = & 4.05 \times 10^{-7} \frac{n_{\odot}}{5.3 \times 10^{-4} pc^{-3}} \int_0^1 \frac{\rho(x)}{\rho_0} x (1 - x) dx
\end{eqnarray}
where $D_s=50$~kpc. From the simulations, the integral $\sim 0.0017$, so that
$\tau = 6.9 \times 10^{-10}$, i.e. a negligible contribution to the observed lensing optical depth $1.2 \times 10^{-7}$.
The essential problem is just that the total mass in the population is very small. A similar estimate for the
DR population of the LMC itself yields $\sim 2.1 \times 10^{-9}$.
So, this population, either in the Galaxy or the LMC, is unlikely to be a viable solution to the
microlensing problem.

\section{Conclusion}

If the majority of Type~Ia supernovae in the Galaxy are the result of
single degenerate binaries with slightly evolved main sequence donors, the post supernova
kinematics of the donors leads to a population of high velocity white dwarfs. These can potentially
explain the interesting observations of Oppenheimer et al. (2001). Not only is the density 
sufficient to explain the observations, but the age distribution is more characteristic of
the thin disk than the thick disk, as is suggested by the Oppenheimer et al. luminosity function
(Hansen 2001).

The observational test of this proposal is to search for evidence of binarity amongst the
high velocity white dwarfs, by obtaining accurate trigonometric parallaxes and searching for
common proper motion main sequence companions, particularly amongst the hotter white dwarfs,
where the contribution of the spheroid or thick disk is smaller.
Although the DR represent a new population of high velocity white dwarfs and might
be considered interesting from the point of view of microlensing, the total mass involved
is quite small and the expected contribution to the microlensing optical depth from this population is
negligible unless the Galaxy retains only a small fraction of the total iron produced by Type~Ia supernovae.

\acknowledgements
B.H.\ acknowledges support by NASA through Hubble Fellowship grant HF--01120.01--99A from the Space
Telescope Science Institute, which is operated by the Association of Universities of Research in
Astronomy, Inc., under NASA contract NAS5--26555. Phil Armitage, Jay Farihi, Vicky Kalogera, Leon Koopmans, Bohdan Paczynski
and an anonymous referee 
provided valuable comments.

\clearpage

\figcaption[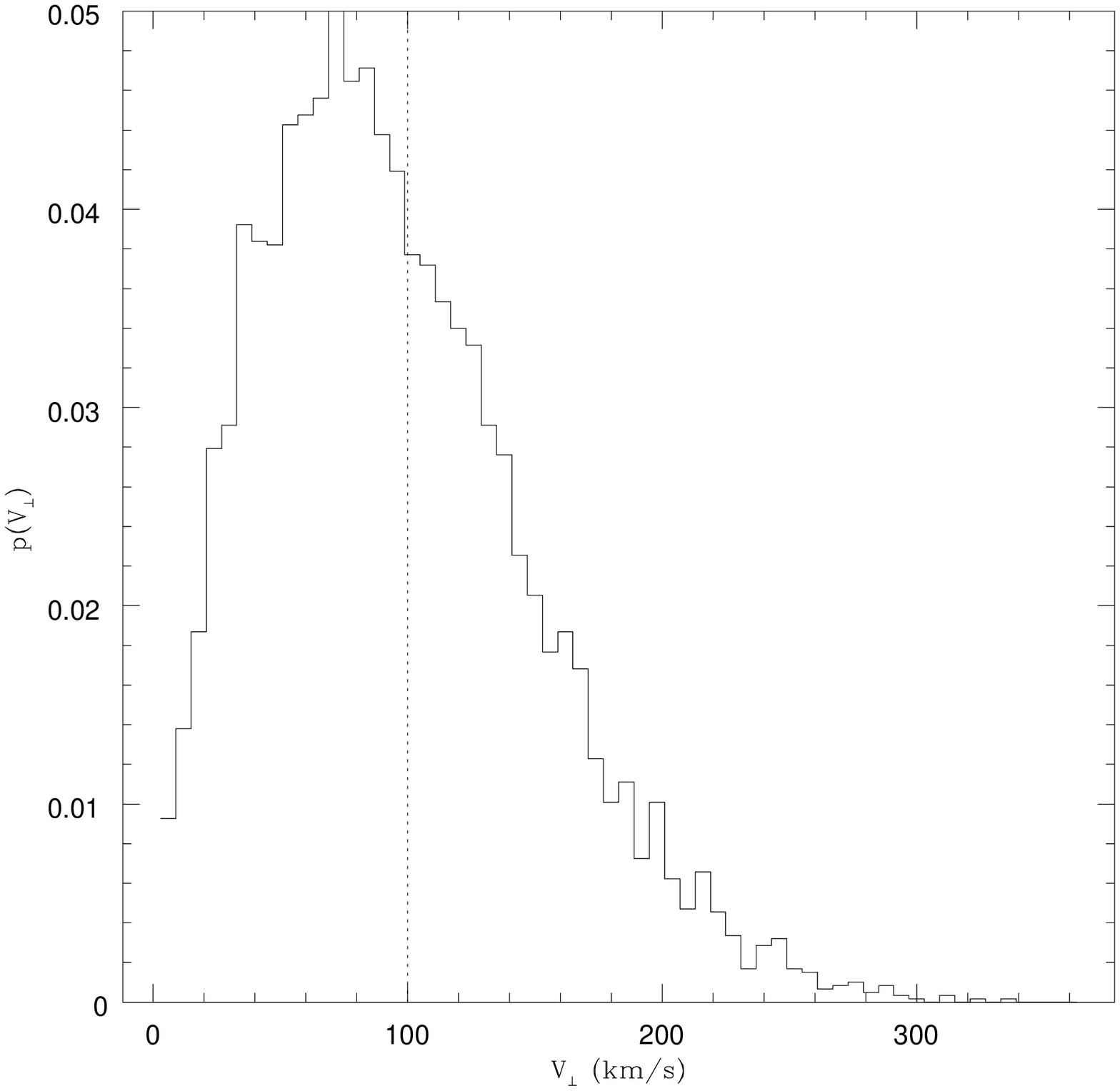]
{The distribution of proper motions for donor remnants within 1~kpc of the sun,
based on the simulation.
 We find that
$34 \%$ of the systems lie above the $100 \rm km.s^{-1}$ cut applied by Oppenheimer et al.
 }

\figcaption[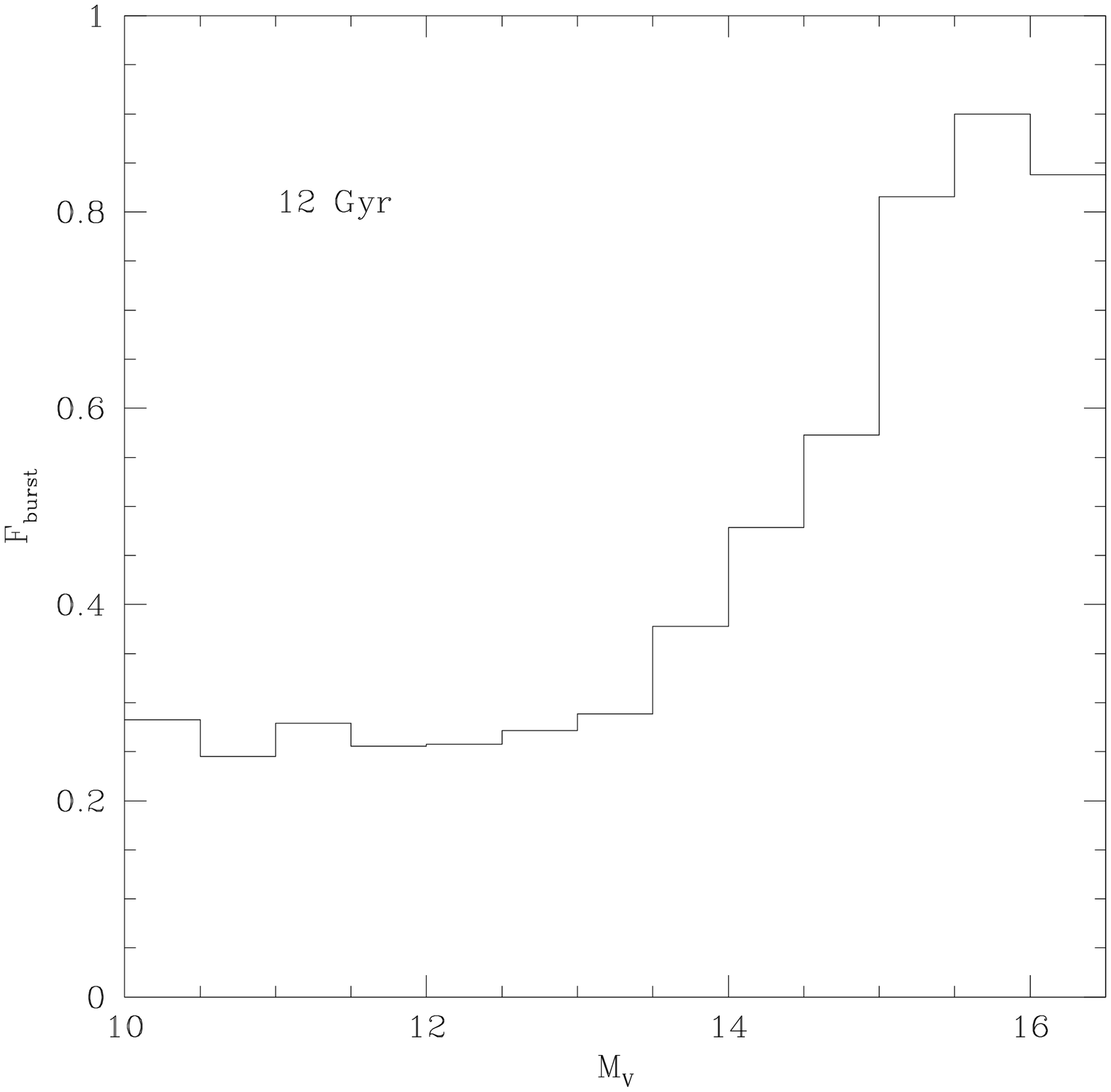]
{ The fraction of stars $N_{\rm burst}/(N_{\rm burst}+N_{\rm DR})$, in any given $M_{\rm V}$ bin, drawn from the burst population.
The total integrated number of stars in both populations is the same and the age of the burst is
assumed to be 12~Gyr. The DR population has been forming stars at a constant rate over the same period. Thus, we see that
the burst stars pile up at the faint end and are less common at the bright end.}

\clearpage
\plotone{f1.ps}
\clearpage
\plotone{f2.ps}

\begin{references}
\reference{Af} Afonso, C. et al., 2000, ApJ, 532, 340
\reference{Alb} Albrow, M. D., et al., 1999, ApJ, 512, 672
\reference{Al01} Alcock, C., et al., 2001, ApJ, 552, 259
\reference{Alc} Alcock, C., et al., 2000, ApJ, 542, 281
\reference{Al97} Alcock, C., et al., 1997, ApJ, 486,697
\reference{Arn} Arnett, W. D., Schramm, D. N. \& Truran, J. W., 1989, ApJ, 339, L25
\reference{BLY} Branch, D., Livio, M., Yungelson, L. R., Boffi, F. R. \& Baron, E., 1995, PASP, 107, 1019
\reference{CMR} Canal, R., Mendez, J. \& Ruiz-Lapuente, P., 2001, ApJ, 550, L53
\reference{DKR} Davies, M. B., King, A. R. \& Ritter, H., 2001, MNRAS in press, /astro-ph/0108106
\reference{DN} Di Stefano, R. \& Nelson, L. A., 1996, in  Lecture Notes in Physics, Vol. 472, Super-Soft X-Ray Sources,
ed. J. Greiner,  (Berlin. Springer-Verlag)
\reference{Han} Hansen, B., 2001, ApJ, 558, L39
\reference{HN} Hillebrandt, W. \& Niemeyer, J. C., 2000, ARA\&A, 38, 191
\reference{HOS} Holberg, J. B., Oswalt, T. D. \& Sion, E. M., 2002, ApJ, 571, 512
\reference{K99} Kahabka, P., 1999, A\&A, 344, 459
\reference{KvdH} Kahabka, P. \& van den Heuvel, E. P. J., 1997, ARA\&A, 35, 69
\reference{KB} Koopmans, L. V. E. \& Blandford, R. D., 2001, /astro-ph/0107358
\reference{LDWH} Langer, N., Deutschmann, A., Wellstein, S. \& H\"{o}flich, 2000, A\&A, 362, 1046
\reference{LRB} Leggett, S. K., Ruiz, M.-T. \& Bergeron, P., 1998, ApJ, 497, 294
\reference{LDM} Liebert, J., Dahn, C. C., \& Monet, D. G., 1988, ApJ, 332, 891
\reference{Opp} Oppenheimer, B. R., Hambly, N. C., Digby, A. P., Hodgkin, S. T. \& Saumon, D., 2001, Science, 292, 698
\reference{Pac} Paczynski, B., 1990, ApJ, 348, 485
\reference{QW} Qian, Y.-Z., \& Wasserburg, G. J., 2002, ApJ, 567, 515
\reference{RDS} Rappaport, S., Di Stefano, R. \& Smith, J. D., 1994, ApJ, 426, 697
\reference{RSH} Reid, I. N., Sahu, K. C. \& Hawley, S. L., 2001, ApJ, 559, 942
\reference{RRC} Reyle, C., Robin, A. C. \& Creze, M., 2001, A\&A, 378, L53
\reference{RC} Ruiz-Lapuente, P. \& Canal, R., 2001, ApJ, submitted
\reference{S94} Sahu, K. C., 1994, Nature, 370, 275
\reference{Tam} Tammann, G. A., Loffler, W. \& Schroder, A., 1994, ApJS, 92, 487
\reference{vdBM} van den Bergh, S. \& McClure, R. D., 1994, ApJ, 425, 205
\reference{YLT} Yungelson, L., Livio, M., Truran, J. W., Tutukov, A. \& Federova, A., 1996, ApJ, 466, 890
\reference{ZE} Zhao, H. \& Evans, N. W., 2000, ApJ, 545, L35

\end{references}
\end{document}